\documentclass[tradiabstract]{aa} 
\usepackage{graphicx}
\usepackage{txfonts}
\usepackage{natbib}
\begin{document}
   \title{Photospheric activity and rotation of the planet-hosting star CoRoT-Exo-4a\thanks{Based on observations obtained with CoRoT, a space project operated by the French Space Agency, CNES, with partecipation of the Science Programme of ESA, ESTEC/RSSD, Austria, Belgium, Brazil, Germany, and Spain.}} 
  
\authorrunning{A. F. Lanza et al.}
\titlerunning{Activity and  rotation in CoRoT-Exo-4a}

   \author{A.~F.~Lanza\inst{1} \and S.~Aigrain\inst{2}
          \and S.~Messina\inst{1} \and G.~Leto\inst{1} \and
          I.~Pagano\inst{1} 
         \and M.~Auvergne\inst{3} \and A.~Baglin\inst{3} 
       \and P.~Barge\inst{4} \and A.~S.~Bonomo\inst{1,4,5} \and A.~Collier Cameron\inst{6} \and G.~Cutispoto\inst{1} \and M.~Deleuil\inst{4}
\and J.~R.~De~Medeiros\inst{7} 
\and B.~Foing\inst{8}  \and C.~Moutou\inst{4}
          }

   \institute{INAF-Osservatorio Astrofisico di Catania, Via S.~Sofia, 78, 95123 Catania, Italy\\
              \email{nuccio.lanza@oact.inaf.it}
         \and
    School of Physics, University of Exeter, Stocker Road, Exeter, EX4 4QL, United Kingdom
\and 
    LESIA, CNRS UMR 8109, Observatoire de Paris, 5 place J. Janssen, 92195 Meudon, France     
   \and  
Laboratoire d'Astrophysique de Marseille (UMR 6110),
Technopole de Ch\^{a}teau-Gombert,
38 rue Fr\'ed\'eric Joliot-Curie,
13388 Marseille cedex 13, France     
   \and 
    Dipartimento di Fisica e Astronomia, Universit\`a degli Studi di Catania, Via S. Sofia, 78, 95123 Catania, Italy     
    \and 
    School of Physics and Astronomy, University of St. Andrews, 
    North Haugh, St Andrews, Fife Scotland KY16 9SS   
    \and 
    Departamento de F\'{\i}sica, Universidade Federal do Rio Grande do Norte, 59072-970 Natal, RN, Brazil
   \and 
   ESA/ESTEC/SRE-S, Postbus 299, 2200 AG Noordwijk, The Netherlands}

   \date{Received ; accepted }

 
  \abstract
   {The space experiment CoRoT ({\it Convection, Rotation and Transits}) has recently detected a transiting hot Jupiter in orbit around a moderately active F-type main-sequence star (CoRoT-Exo-4a). This planetary system is of particular interest because it has {{ an orbital period of 9.202 days, the second longest one}} among the transiting planets known to date.}
   {We study the surface rotation and the activity of the host star during an uninterrupted sequence of optical observations of 58 days.}
   {Our approach is based on a maximum entropy spot modelling technique  extensively tested by modelling the variation of the total solar irradiance. Recently, it  has been successfully applied to model the light curve of another active star with a transiting planet observed by CoRoT, i.e., CoRoT-Exo-2a. It assumes that stellar active regions consist of cool spots and bright faculae, analogous to sunspots and solar photospheric faculae, whose visibility is modulated by stellar rotation. }
   { The modelling of the light curve of CoRoT-Exo-4a reveals three main active longitudes with lifetimes between $\sim 30$ and $\sim 60$ days that rotate {{ quasi-synchronously}} with the orbital motion of the planet. The different rotation rates of the  active longitudes are interpreted in terms of surface differential rotation and a lower limit of $0.057 \pm 0.015$ is derived for its relative amplitude. The enhancement of activity observed {{ close to}} the subplanetary longitude suggests a magnetic star-planet interaction, although the short duration of the time series prevents us from drawing definite conclusions.    } 
  {The present work confirms the {{ quasi-synchronicity }} between stellar rotation and planetary orbital motion in the CoRoT-Exo-4 system and provides for the first time a lower limit for the surface differential rotation of the star.  This information can be important {{  in trying to understand}} the formation and evolution of this highly interesting planetary system. Moreover, there is an indication for a possible star-planet magnetic interaction that needs to be confirmed by future studies.}

\keywords{stars: magnetic fields -- stars: late-type -- stars: activity -- stars: rotation -- planetary systems -- stars: individual (CoRoT-Exo-4a)}

   \maketitle
%

\section{Introduction}

CoRoT is a photometric space experiment devoted to asteroseismology and the search for extrasolar planets by the method of transits \citep{Baglinetal06}. It has recently discovered \object{CoRoT-Exo-4b}, a Jupiter-sized planet transiting across the disc of an F-type main-sequence star with an orbital period of 9.202 days 
\citep{Aigrainetal08}. This is the second longest period among the transiting planetary systems known to date, putting CoRoT-Exo-4b in a region of the mass-period parameter space that was previously empty \citep[cf.][]{Moutouetal08}. Moreover,  the out-of-transit light curve shows a modulation with an amplitude of a few 0.001 mag that can be attributed to photospheric brightness inhomogeneities carried into and out of view by the rotation of the star. Given its late  spectral type, those inhomogeneities can be considered analogous to cool spots and bright faculae observed in the Sun, owing their existence to photospheric magnetic fields. 

The active regions of CoRoT-Exo-4a are sufficiently stable to allow an estimate of its rotation period through an autocorrelation analysis, yielding a period of $8.87 \pm 1.12$ days. This result indicates that the stellar rotation and the orbital motion of the planet are  {{ quasi-synchronized}}, which cannot be explained on the basis of the tidal theory given the large separation between the star and the planet ($a/R_{\rm star} = 17.36 \pm 0.25$) and its low mass \citep[$0.72 \pm 0.08$ M$_{\rm Jup}$, cf., ][]{Aigrainetal08}.  Other synchronous or quasi-synchronous star-planet systems have recently been found, consisting of quite massive planets orbiting F-type stars \citep{Catalaetal07,McCulloughetal08}. Among those systems, \object{$\tau$ Bootis}
has been recently studied in some detail revealing a  surface differential rotation with a relative amplitude  $\Delta \Omega / \Omega \approx 0.15-0.18$ and hints for an activity cycle of  a few years \citep{Catalaetal07,Donatietal08}. 
However, those systems host more massive and closer planets that may explain the synchronization on the basis of tidal effects, assuming that only the outer stellar convection zone, having a mass of only a few 0.001 of the stellar mass, is synchronized \citep[cf. ][]{Donatietal08}. { On the other hand, the estimated time scale for the synchronization of the outer convective envelope of CoRoT-Exo-4a is $\approx 350$ Gyr, assuming a mass of 0.001  of the  stellar mass for the envelope.}

In the present work, we present a modelling of the out-of-transit light curve of CoRoT-Exo-4a following the approach  already applied for another star hosting a transiting hot Jupiter, i.e., \object{CoRoT-Exo-2a}. It is based on a solar analogy, i.e., it assumes that the stellar active regions responsible for the flux modulation consist of cool spots and bright faculae \citep{Lanzaetal08}.

Late-type stars accompanied by hot Jupiters sometimes show active regions rotating with the orbital period of the planet instead of the stellar rotation period. The phenomenon has been detected in some seasons in \object{$\upsilon$ Andromedae} and \object{HD~179949} that have shown chromospheric hot spots leading the planet by $\sim 170^{\circ}$ and
 $\sim 70^{\circ}$, respectively \citep{Shkolniketal08}. Moreover, the synchronous system \object{$\tau$ Boo} has shown some evidence for  
 an active region rotating with the orbital period of its planet but leading it by $\approx 70^{\circ}$ during the period 2001-2005 \citep{Walkeretal08}. These phenomena suggest a magnetic interaction between the star and its hot Jupiter, as discussed by \citet{Lanza08}. In the case of CoRoT-Exo-4a, the short duration of the observations prevents us from deriving definite conclusions about a possible star-planet interaction, but the longitude distributions of the active regions may be compared with the orbital position of the planet to see whether there is some analogy, say,  with $\tau$ Boo.

A detailed study of stellar activity by means of Doppler Imaging techniques is not feasible owing to the small projected rotation velocity of CoRoT-Exo-4a ($v \sin i = 6.4 \pm 1.0$ km s$^{-1}$) and its faintness ($R \sim 13.5$). Therefore, a spot modelling approach is the most suitable method to investigate the magnetic activity and rotation of this very interesting object. 

\section{Observations}

CoRoT-Exo-4a was observed during the CoRoT Initial Run for 58 days, starting from 6 February 2007. The time sampling was initially of 512 s, then reduced to 32 s during the last three weeks after the detection of the transits by  the so-called CoRoT alarm-mode. CoRoT performed aperture photometry with a fixed mask. Since the star was brighter than $R = 14.5$, the flux was split along detector column boundaries into broad-band red, green and blue channels.   

The observations and data processing are described by \citet{Aigrainetal08} to whom we refer the reader for details. The reduction pipeline applied corrections for the background and the pointing jitter of the satellite which was particularly relevant during  ingress and  engress from the Earth shadow. Measurements during the crossing of the South Atlantic Anomaly of the Earth magnetic field, which amounted to about $15-20$ percent of a satellite orbit, were discarded. More information on the instrument, its operation and performance can be found in \citet{Auvergneetal08}. 

The light curve extracted from the CoRoT Mission Archive\footnote{http://idoc-corot.ias.u-psud.fr/} was further corrected for the effect of a hot pixel in the blue channel, as explained by \citet{Aigrainetal08}. To increase the signal-to-noise ratio and reduce residual systematic effects possibly present in individual channels, we summed up the flux in the red, green and blue channels to obtain a  light curve in a spectral range extending from 300 to 1100 nm. The oversampled section of this white-band light curve  was rebinned with a regular 512 s sampling, to obtain an evenly sampled time series. It was further cleaned by applying a moving median $5$-$\sigma$ clipping filter that allowed us to identify and discard residual outliers, as described by  \citet{Aigrainetal08}. The standard deviation of the observations, averaged over 512~s bins, is $8.9 \times 10^{-4}$ in relative flux units. 
The contamination of the photometric aperture by stars other than CoRoT-Exo-4 was only $0.3 \pm 0.1$ percent of the median of the measured flux and it was subtracted to avoid dilution of the stellar light variations. Transits were removed by means of the ephemeris and parameters of \citet{Aigrainetal08}. 

The flux variations related to stellar activity have time scales of the order of a day or longer. Therefore, we  rebinned the out-of-transit light curve by computing normal points along each orbital period of the satellite (6184 s). This has the advantage of removing tiny systematic variations that may still be associated with the orbital motion of the satellite \citep[cf. ][]{Alonsoetal08,Auvergneetal08}. Since the orbital period is not a multiple of the individual exposures of 32 or 512 s, we performed a linear interpolation of the flux variation along the last 40 s of each orbital period to compute the normal points. In such a way we obtained a  light curve consisting of 684 normal points ranging from HJD 2454135.0918 to HJD 2454192.7744, i.e., with a duration of 57.6826 days.  

The light curve shows a long-term decreasing trend \citep[cf. Fig.~2 in ][]{Aigrainetal08} that may be of instrumental origin \citep{Auvergneetal08}. Since it would produce only a monotonous increase of the spotted area in our model along the $\sim 60$ days of the present observations, we decided to subtract it before modelling the light curve by fitting a straight line to the data. Finally, the de-trended light curve  was normalized at its maximum flux value observed at 
HJD~$=2455737.2730$, that we assumed to represent the unspotted flux level of the star, whose true value is unknown.

\section{Spot modelling}
\label{spotmodel}

{The reconstruction of the surface brightness distribution from the rotational modulation of the stellar flux is an ill-posed problem, because the variation of the flux vs. rotational phase contains information only on the distribution of the brightness inhomogeneities vs. longitude. The integration over the stellar disc cancels any latitudinal information, particularly when the inclination of the rotation axis along the line of sight is $90^{\circ}$, as in the present case \citep[see Sect.~\ref{model_param} and ][]{Lanzaetal08}. Therefore, we need to include a priori information in the light curve inversion process to obtain a unique and stable map. This is  done by computing a Maximum Entropy (hereinafter ME) map which has been proved to give the best representation in the case of the Sun \citep[cf. ][]{Lanzaetal07}. As a matter of fact, a sequence of seven maps is obtained for CoRoT-Exo-4a, covering its successive rotations along the 58 days of the observations. They will allow us to study the evolution of the active longitudes over the surface of the star.
Readers not interested in the details of our spot modelling approach may skip the remainder of this Section and go on to Sect.~\ref{model_param} (or directly to Sect.~\ref{results}) }. 

In our model the star is subdivided into a large number of surface elements, in our case 200  squares of side $18^{\circ}$, with  each element containing unperturbed photosphere, cool spots and facular areas. The fraction of an element covered by cool spots is indicated by the filling factor $f$,  the fractional  area of the faculae is $Qf$ and the fractional area of the unperturbed photosphere is $1-(Q+1)f$. We fit the light curve by changing the value of $f$ over the surface of the star, while $Q$ is held uniform and constant. Even fixing the rotation period, the inclination, and the spot and facular contrasts \citep[see ][ for details]{Lanzaetal07}, the model has 200 free parameters and suffers from  non-uniqueness and instability. To find a unique and stable spot map, we apply maximum entropy regularization, as described in \citet{Lanzaetal07}, by minimizing a functional 
$\Theta$ which is a linear combination of the $\chi^{2}$ and  the entropy functional $S$, i.e.:
\begin{equation}
\Theta = \chi^{2} ({\vec f}) - \lambda S ({\vec f}),
\end{equation}
where ${\vec f}$ is the vector of the filling factors of the surface elements, $\lambda > 0$  a Lagrangian multiplier determining the trade-off between light curve fitting and regularization, and the expression of $S$ is given in \citet{Lanzaetal98}. { The entropy functional $S$ is constructed in such a way that it attains its maximum value when the star is immaculate. Therefore, by increasing the Lagrangian multiplier $\lambda$, we increase the weight of $S$ in the model and the area of the spots  is progressively reduced.
This gives rise to systematically negative residuals between the observations and the best fit model when
$\lambda > 0$. The optimal value of $\lambda$ depends on the information content of the light curve, that in turn depends on the ratio of the amplitude of its rotational modulation to the average standard deviation of its normal points. } To optimize the extraction of this information, we generalize the criterium to fix the Lagrangian multiplier previously introduced in \citet{Lanzaetal08}. 
For a given value of $\lambda$, we compute the mean of the residuals for the regularized best fit $\mu_{\rm reg}$ and compare it with $\epsilon_{0} \equiv \sigma_{0}/\sqrt{N}$, i.e., the standard error of the residuals, where $\sigma_{0}$ is their standard deviation,  obtained in the case of the unregularized best fit (i.e., for $\lambda=0$), and $N$ is the number of normal points in each fitted subset of the light curve of duration  $\Delta t_{\rm f}$ (see below).  
 We iterate on the value of $\lambda$ until $|\mu_{\rm reg}| = \beta \epsilon_{0}$, where $\beta$ is a   numerical factor that { will be fixed a posteriori according to the requisites of an acceptable fit and a regular overall evolution of the spots (cf. Sect.~\ref{results}). } In the case of the light curve of CoRoT-Exo-2a, the rotational modulation has an amplitude of $\sim 0.06$ mag, which corresponds to a signal-to-noise ratio of $\sim 300$ for a standard deviation of $\approx 2 \times 10^{-4}$ mag. In the present case, the amplitude of the rotational modulation of CoRoT-Exo-4a is only $\sim 0.006$ mag and the standard deviation of the points is somewhat greater because the star is fainter, giving a signal-to-noise ratio of $\sim 20$. Therefore, while $\beta \simeq 1$ was found to be adequate for CoRoT-Exo-2a, $\beta > 1$ is required to appropriately reconstruct ME maps of CoRoT-Exo-4a.   

In the case of the Sun, assuming a fixed distribution of the filling factor, it is possible to obtain a good fit of the irradiance changes only for a limited time interval $\Delta t_{\rm f}$, not exceeding 14 days that is the lifetime of the largest sunspot groups dominating the irradiance variation. In the case of other active stars, the value of        $\Delta t_{\rm f}$ must be determined from the observations themselves, looking for the maximum data extension that allows for a good fit with the applied model (see Sect.~\ref{model_param} for CoRoT-Exo-4a). 

The optimal values of the spot and facular contrasts, and of the facular-to-spotted area ratio $Q$ in stellar active regions are  unknown a priori. In our model the facular contrast $c_{\rm f}$ and the parameter $Q$ enter as the product $c_{\rm f} Q$, hence we can fix $c_{\rm f}$ and vary $Q$, estimating its best value 
 by minimizing the $\chi^{2}$ of the model, as will be shown in Sect.~\ref{model_param}. Since the number of free parameters of the ME model is large, for this specific application we shall make use of the model of \citet{Lanzaetal03} which fits the light curve by assuming only three active regions to model the rotational modulation of the flux plus a uniformly distributed background to account for the variations of the mean light level. This procedure is the same adopted to fix the value of $Q$ in the case of CoRoT-Exo-2a 
\citep[cf. ][]{Lanzaetal08}.  

{ As in the case of CoRoT-Exo-2a, we shall assume an inclination of the rotation axis of CoRoT-Exo-4a of $90^{\circ}$,  which implies that our model cannot provide information on the latitude distribution of the active regions. Such a limitation cannot be overcome by assuming an inclination $ i < 90^{\circ}$ in the model, owing to the relatively low information content of the present light curve. Specifically, when we assume $ i < 90^{\circ}$, we find that the ME regularization virtually puts all the spots at the sub-observer latitude (i.e., $90^{\circ} -i$) to minimize their area and maximize the entropy. Therefore, we are limited to map only the distribution of the active regions vs. longitude, that can be done with a resolution of at least   $\sim 50^{\circ}$ \citep[cf. ][]{Lanzaetal07,Lanzaetal08}.} Note that our ignorance of the true value of the facular contribution to the light variations may produce systematic errors in the active region longitudes derived by our model, as discussed  by \citet{Lanzaetal07} in the case of the Sun.

\section{Model parameters}
\label{model_param}

The fundamental stellar parameters are taken from \citet{Aigrainetal08} and \citet{Moutouetal08} and are listed in 
Table~\ref{model_param_table}. A quadratic limb-darkening law is adopted  for the stellar photosphere, viz., 
$I(\mu) \propto (a_{\rm p} + b_{\rm p}\mu + c_{\rm p}\mu^{2})$, where $I(\mu)$ is the specific intensity at the limb position $\mu \equiv \cos \theta$, with $\theta$ being the angle between the normal to a surface element and the line of sight \citep[cf. ][]{Lanzaetal03,Lanzaetal07}. The limb-darkening parameters have been derived from \citet{Kurucz00} model atmospheres for $T_{\rm eff} = 6190$ K, $\log g = 4.41$ (cm s$^{-2}$) and solar abundances, adopting the CoRoT white-band transmission profile in \citet{Auvergneetal08}. 

\begin{figure}[t]
\centerline{
\includegraphics[width=8cm,height=6cm]{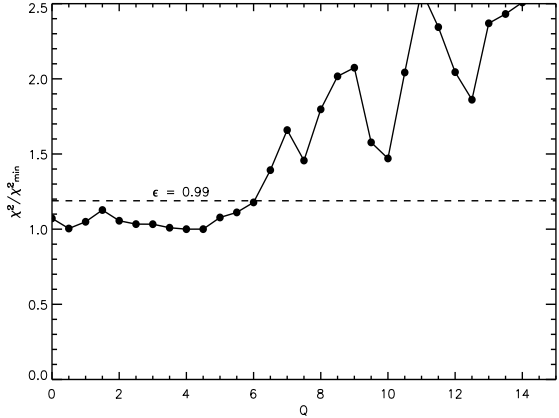}} 
\caption{The ratio of the $\chi^{2}$ of the composite best fit to the entire time series of CoRoT-Exo-4a
to its minimum value vs. the parameter $Q$, i.e., the ratio of the area of the faculae to that of the cool spots in active regions. The horizontal dashed line indicates the 99 percent confidence level for $\chi^{2}/\chi_{\rm min}^{2}$, determining the interval of the acceptable $Q$ values.}
\label{qratio}
\end{figure}
\begin{table}
\noindent 
\caption{Parameters adopted for the spot modelling of CoRoT-Exo-4a.}
\begin{tabular}{lrr}
\hline
 & & \\
Parameter &  & Ref.$^{a}$\\
 & & \\ 
\hline
 & &  \\
Star Mass ($M_{\odot}$) & 1.16 & A08  \\
Star Radius ($R_{\odot}$) & 1.17 & A08  \\
$T_{\rm eff}$ (K) & 6190 &  M08 \\ 
$a_{\rm p}$ & 0.334 & L09 \\
$b_{\rm p}$ & 1.032 & L09 \\
$c_{\rm p}$ & -0.381 & L09 \\ 
$P_{\rm rot}$ (d) & 9.20205 & A08 \\
$\epsilon$ & $1.10 \times 10^{-4}$ & L09 \\ 
Inclination (deg) & 90.0 & A08  \\
$c_{\rm s}$  & 0.681 & L09 \\
$c_{\rm f}$  & 0.115 & L04 \\ 
$Q$ & 4.5  & L08 \\ 
$\Delta t_{\rm f}$ (d) & 8.24037 & L09 \\ 
& &   \\
\hline
\label{model_param_table}
\end{tabular}
~\\
$^{a}$ References: A08: Aigrain al. (2008); M08: Moutou et al. (2008); L04: Lanza et al. (2004); L09: present study.
\end{table}

The polar flattening of the star due to the centrifugal potential is computed in the Roche approximation with a rotation period of 9.202 days { (see below)}. The relative difference between the equatorial and the polar radii is $\epsilon = 1.10 \times 
10^{-4}$ which may induce a flux variation about two orders of magnitude smaller for a spot coverage of $\sim 1$ percent as a consequence of the gravity darkening of the equatorial regions of the star. 

The inclination of the stellar rotation axis is difficult to constrain through the observation of the Rossiter-McLaughlin effect because of the small $ v \sin i$ of the star and its intrinsic line profile variations due to stellar activity \citep{Moutouetal08}. Nevertheless, we shall assume that the stellar rotation axis is normal to the orbital plane of the planet, i.e., at an inclination of $90^{\circ}$ from the line of sight \citep[cf. ][]{Aigrainetal08}. 

The rotation period adopted for our spot modelling is equal to the orbital period of the planet. This will allow us to check the synchronization of stellar rotation to the planetary orbit through the study of the longitude drift of the active regions versus time (cf. Sect.~\ref{results}). { Another advantage of this choice is that the subplanetary longitude is fixed in our reference frame, which allows us to investigate straightforwardly the possible association between active longitudes and orbital position of the planet. Nevertheless, considering that the autocorrelation analysis of the light modulation gives a rotation period of $8.87 \pm 1.12$ days \citep{Aigrainetal08}, it is worth investigating also spot models with a rotation period of 8.87 days.} 

The maximum time interval that our model can accurately fit with a fixed distribution of active regions, has been determined by means of a series of tests and has been found to be of $\sim 8.2$ days. Therefore, we subdivided the time series into  seven equal segments of duration $\Delta t_{\rm f} = 8.24037$ days, each of which  was modelled with a fixed active region pattern. 

To compute the spot contrast, we adopted the same mean temperature difference derived in the case of sunspot groups from their bolometric contrast, i.e. 560 K \citep{Chapmanetal94}. In other words, we assumed a spot effective temperature of $ 5630$ K, yielding a constrast $c_{\rm s} = 0.681$ in the CoRoT passband \citep[cf. ][]{Lanzaetal07}.
A different spot contrast changes the absolute spot coverages, but does not significantly affect their longitudes and their time evolution, as discussed in detail by \citet{Lanzaetal08}. The facular contrast was assumed to be solar-like with $c_{\rm f} = 0.115$ \citep{Lanzaetal04}. 

The best value of the area ratio $Q$ between the faculae and the spots in each active region has been estimated by means of the model by \citet[][ cf. Sect.~\ref{spotmodel}]{Lanzaetal03}. In Fig.~\ref{qratio}, we plot the ratio $\chi^{2}/ \chi^{2}_{\rm min}$ of the 
total $\chi^{2}$ of the composite best fit to the entire time series to 
its minimum value $\chi^{2}_{\rm min}$, versus $Q$. The horizontal dashed line indicates the 99 percent confidence level as derived from the F-statistics \citep[e.g., ][]{Lamptonetal76}. The best value of $Q$ turns out to be $Q=4.5$, with an acceptable range extending from $\sim 0$ to $\sim 6.5$. For comparison, the best value in the case of the Sun is $Q_{\odot}=9$, indicating a lower relative contribution of the faculae to the total light variation in CoRoT-Exo-4a. 
The amplitude of the rotational modulation of our star is $\sim 0.006$ mag, i.e., $2-3$ times that of the Sun at the maximum of the eleven-year cycle  (see Sect.~\ref{results}). This suggests that CoRoT-Exo-4a is more active than the Sun, which may account for the reduced facular contribution to its light variations, as suggested by \citet{Radicketal98} and \citet{Lockwoodetal07}.  
\begin{figure*}[]
\centerline{
\includegraphics[width=12cm,angle=90]{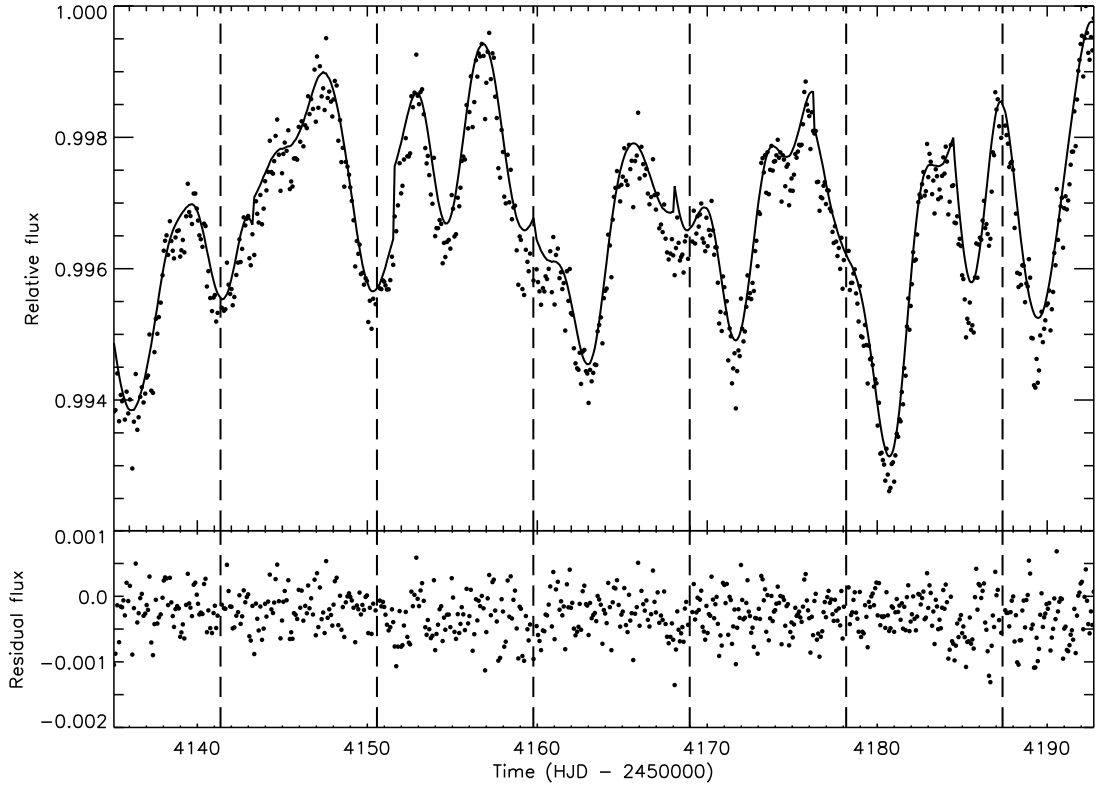}} 
\caption{{\it Upper panel:} The ME-regularized composite best fit to the out-of-transit light curve of CoRoT-Exo-4a obtained for $Q=4.5$. The flux is in relative units, i.e., measured with respect to the maximum observed flux in the de-trended normal point light curve. {\it Lower panel:} The residuals from the composite best fit versus time. { In both panels, the dashed vertical lines indicate the epochs of mid transits.}   
}
\label{bestfit}
\end{figure*}

\section{Results}
\label{results}

The composite best fit obtained with the ME regularization is shown in Fig.~\ref{bestfit}, together with the residuals. { The systematic negative values of the residuals are a consequence of the ME regularization that tends to reduce the spotted area as much as possible, thus systematically driving the fit above the observations (cf. Sect.~\ref{spotmodel}).}

The residuals of the unregularized composite best fit have a mean of $7.845 \times 10^{-7}$ and a standard deviation  
$\sigma_{0} = 2.832 \times 10^{-4}$ in relative flux units, i.e., assuming the maximum observed flux as the unit of measure. 
{ Note that  in the absence of regularization the standard deviation of the residuals is larger than expected from the photon shot noise, i.e., $1.61 \times 10^{-4}$ \citep{Aigrainetal08}. This may arise from intrinsic stellar microvariability on short-term time scales (up to a few days) or may be due to some residual instrumental effects. }
On the average, we have $N \sim 100$ normal points per fitted subset. The Lagrangian multipliers for the regularized ME models have been fixed to obtain a mean of the residuals of the composite fit $ \mu_{\rm reg }= -2.731 \times 10^{-4}$, corresponding to $|\mu_{\rm reg}| \simeq 9.6 \, \epsilon_{0}$.
The optimal value $\beta = 9.6$ was established by considering the information content of the maps obtained for different values of $\beta$ and minimizing it (i.e., maximizing the entropy) as much as possible, while still retaining an average acceptable fit. Smaller values of $\beta$ give an improvement of the fit over small sections of the light curve, but introduce several small spots that vary from one subset to the other. Therefore, they are  actually not required by the data. On the other hand, by increasing the entropy of the maps as much as possible, we obtain smoother maps with a regular variation of the spot pattern from one subset to the other, which indicates that we are properly  modelling the overall active region distribution on the star.  The two spikes of the best fit at  HJD $\sim 2454160$ and $\sim 2454168$ are due to the small variations of the spot configuration when we move from one data subset to the next. { The reader may wonder about the possibility of improving the regularized best fit by considering smaller time intervals $\Delta t_{\rm f}$. However, the improvement  is negligible because the misfits at the matching points between one subset and the next are a consequence of the significant amount of regularization required by this light curve due to its relatively low information content. The effect is actually much larger in the present case than for CoRoT-Exo-2a, whose light curve had a significantly greater information content than the present one \citep[cf. ][]{Lanzaetal08}. Therefore, decreasing $\Delta t_{\rm f}$ in the present case does not alleviate the problem, but produces the opposite effect, owing to a greater number of  matching points along the time series. } 

\begin{figure}
\centerline{
\includegraphics[width=7cm]{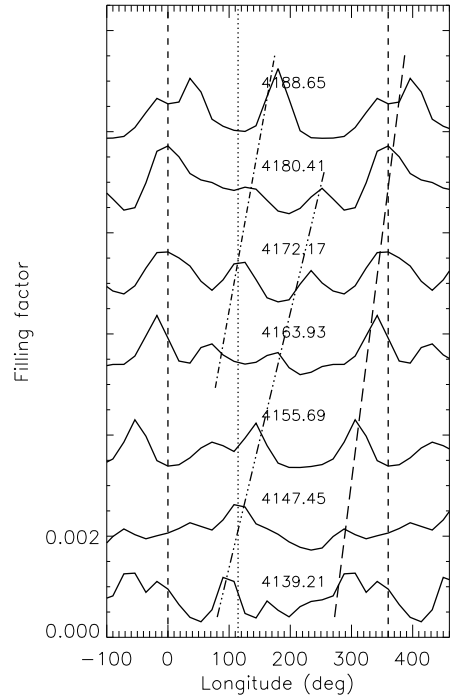}} 
\caption{The distributions of the spotted area vs. longitude at the labelled times (${\rm HJD}-2450000.0$) for $Q=4.5$. The plots have been vertically shifted to show the migration of individual spots (relative maxima of the distributions) versus time. The vertical dashed lines mark longitudes $0^{\circ}$ and $360^{\circ}$, beyond which the distributions have been repeated to help  following the  spot migration. The vertical dotted line marks the subplanetary longitude. The dot-dashed, three-dot-dashed and long-dashed lines trace the migration of the most conspicuous spots detected in the plots (see the text for details). 
}
\label{long_distr}
\end{figure}

The distributions of the spotted area vs. longitude are plotted in Fig.~\ref{long_distr}
for the seven mean epochs of our individual subsets { adopting a rotation period of 9.20205 days}. The longitude zero corresponds to the point intercepted on the stellar photosphere by the line of sight to the centre of the star  at HJD~2454135.0918, i.e., the sub-observer point at the initial epoch. The longitude increases in the same sense as the stellar rotation and the orbital motion of the planet, which has a fixed subplanetary point on the star in the adopted reference frame. This choice allows a direct comparison of the active region longitudes with the subplanetary longitude, although { one cannot correlate the  active regions to the dips in the light curves in a straightforward way.}

{Three main active regions can be identified in Fig.~\ref{long_distr} and their migration has been traced with different straight lines. Specifically, their rotation rates  with respect to the adopted reference frame are found by a linear best fit, i.e., assuming a constant migration rate. They are  listed in Table~\ref{AR_migration} for three values of the facular-to-spotted area ratio $Q$, because the longitudes of the active regions derived from our model have some dependence on that parameter \citep[cf. ][]{Lanzaetal07,Lanzaetal08}. The active region indicated as AR$_{1}$ corresponds to the one marked with a three-dot-dashed line in Fig.~\ref{long_distr}, AR$_{2}$ to that marked with 
a long-dashed line, and AR$_{3}$ to that marked with a dot-dashed line, respectively. 
\begin{table}
\noindent 
\caption{Relative migration rates $\Delta \Omega / \Omega$ of the active longitudes for different facular-to-spotted area ratio $Q$ and rotation period $P_{\rm rot}$ .}
\begin{center}
\begin{tabular}{ccccc}
\hline
 & & &  & \\
$Q$ & $P_{\rm rot}$ & AR$_{1}$ & AR$_{2}$ & AR$_{3}$ \\
& (day)  &  &  &  \\
 & & & \\ 
\hline
 & & &  & \\ 
0.0  & 9.202 & $0.106 \pm 0.025$ & $0.034 \pm 0.008$  &  $0.070 \pm 0.024$ \\ 
4.5  & 9.202 & $0.108 \pm 0.008$ & $0.052 \pm 0.009$ & $0.100 \pm 0.024$ \\
 4.5  & 8.870 & $0.063 \pm 0.011$ & $0.010 \pm 0.010$ & $0.073 \pm 0.036$ \\
7.0 & 9.202 &  $0.100 \pm 0.011$ & $0.046 \pm 0.010$ & $0.128 \pm 0.010$ \\
& & &  &  \\
\hline
\label{AR_migration}
\end{tabular}
\end{center}
\end{table}

The relative amplitude of the surface differential rotation, estimated from the difference between the greatest and the lowest migration rates, is $ \Delta \Omega /\Omega = 0.072 \pm 0.026$, $0.057 \pm 0.015$, and $0.0825 \pm 0.014$ for 
$Q=0$, 4.5, and 7.0, respectively. { For the sake of completeness, we have also computed ME spot models adopting $Q=4.5$ and a rotation period of 8.870 days, that comes from the autocorrelation analysis of the light curve \citep{Aigrainetal08}. Again, three main active longitudes are seen in the longitude distribution of the spotted area. Their migration rates are listed in Table~\ref{AR_migration} and the derived relative amplitude for the differential rotation is 
$\Delta \Omega / \Omega = 0.063 \pm 0.038$, in very good agreement with the result obtained for a rotation period of 9.20205 days. It is interesting to note that a rotation period of 8.870 days makes the migration rate of AR$_{2}$, i.e., the active longitude with the greatest spotted area, virtually zero. This is in agreement with the fact that the autocorrelation of the light curve is maximized for that rotation period.  }

Unfortunately, no information is available on the spot latitudes  in our models, therefore, our $\Delta \Omega / \Omega$ are only  lower limits. These values suggest that CoRoT-Exo-4a has a  surface differential rotation  comparable to that of the Sun, for which we can estimate a relative amplitude of $0.04-0.05$ considering active regions confined within the sunspot belt, i.e., within $\pm \, (35^{\circ} - 40^{\circ})$ from the equator.} 

The lifetimes of the active regions can be estimated from our spot models and range from 10 to 50 days, with longer lifetimes characterizing those having greater filling factors, as in the Sun. 

\begin{figure}
\centerline{
\includegraphics[width=7cm]{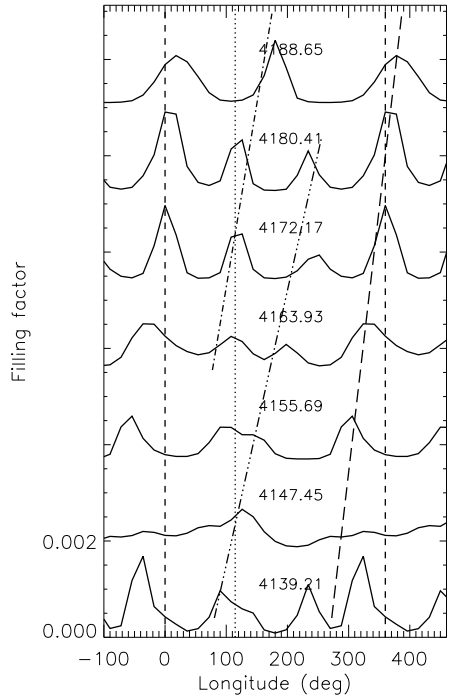}} 
\caption{The same as Fig.~\ref{long_distr}, but for $Q=0.0$.  
}
\label{long_distr0}
\end{figure}

\begin{figure}
\centerline{
\includegraphics[width=7cm,angle=90]{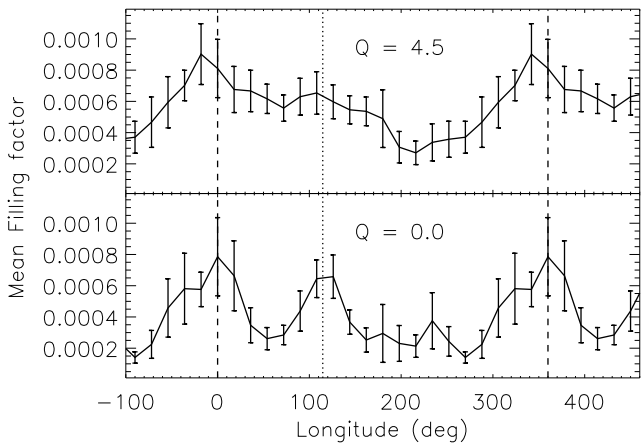}} 
\caption{ {\it Upper panel:} The spotted area per $18^{\circ}$ longitude bin averaged over time versus longitude for $Q=4.5$. The semiamplitudes of the error bars are equal to one standard deviation of the spotted area in the corresponding bin. {\it Lower panel:} The same as in the upper panel, but for $Q=0$. The vertical dotted line marks the subplanetary longitude, while the dashed vertical lines mark longitudes $0^{\circ}$ and $360^{\circ}$ beyond which the distributions are repeated.    
}
\label{long_mean}
\end{figure}

It is  interesting to note that some active regions are located close to the { subplanetary longitude} during most of the time, although the highest concentrations of spotted (and facular) areas are around a longitude following the planet by about $100^{\circ}-120^{\circ}$. Conversely, the hemisphere centred around longitude $\approx 240^{\circ}$ has the minimum active region coverage (cf. Fig.~\ref{long_distr}). 
{The association between the subplanetary longitude and an enhanced spot activity is particularly remarkable in the case of the models computed without facular contribution, i.e., for $Q=0$ (see Fig.~\ref{long_distr0}). In six out of seven epochs, the ME models indicate a relative maximum of the spotted area within $\pm 50^{\circ}$ from the subplanetary longitude. The time average of the spotted area in each longitude bin is plotted in Fig.~\ref{long_mean} vs. longitude, together with the corresponding standard deviation. In the case with $Q=0$ (Fig.~\ref{long_mean}, lower panel), a well-defined relative maximum appears close to the subplanetary longitude, with a height of $\sim 3$ standard deviations, which is unlikely to be due to a chance fluctuation. Moreover, the subplanetary peak has a FWHM significantly smaller than that of the active longitude peak  around $0^{\circ}$ whose width reflects the migration of its active regions along time.  A greater migration rate of the active regions is also responsible for the smearing of the active longitude around $0^{\circ}$ and the subplanetary longitude, when we consider the average of the models computed with $Q=4.5$ (Fig.~\ref{long_mean}, upper panel). }

\begin{figure}[b]
\centerline{
\includegraphics[width=5cm,angle=90]{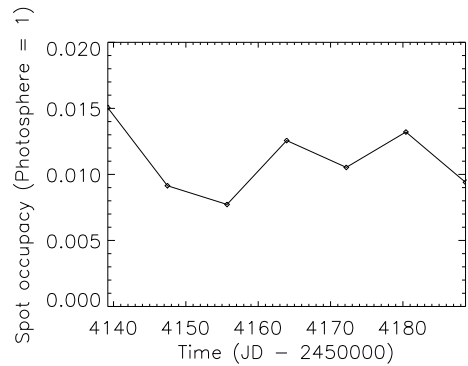}} 
\caption{The total spotted area as derived from the regularized ME models vs. time for $Q=4.5$. 
}
\label{total_area}
\end{figure}

The total spotted area varies only slightly versus time (see Fig.~\ref{total_area}) and may indicate a level of activity about $2-3$ times that of the Sun at the maximum of the eleven-year cycle. It is important to notice that the total spotted area is determined here after removing the long-term decrease of the flux observed in the original light curve by \citet{Aigrainetal08}. Therefore, it does not show the long-term linear increase that one would expect in the case that the long-term trend were not removed. Moreover, the absolute value of the area depends on the adopted spot contrast $c_{\rm s}$ and the value of $Q$. For instance, a lower spot temperature would imply a stronger contrast and thus a smaller area, but the relative variations of the area are largely independent of $c_{\rm s}$ and $Q$ \citep[cf. ][]{Lanzaetal08}.


\section{Discussion}

{ Our results confirm that the rotation of CoRoT-Exo-4a is { quasi-synchronized} with the orbital motion of its planet.  Minimum spot migration rates are obtained for a rotation period of 8.87 days, which  corresponds to the rotation period of the active longitude with the largest spot concentration. However, even adopting that period, the other active longitudes continue to display a significant migration. Our interpretation for the different migration rates of the active longitudes is that the corresponding active regions are located at different latitudes on a differentially rotating star. Nevertheless, the different drift rates of the active longitudes may be produced by a phenomenon different from surface differential rotation. For instance, in the framework of a solar analogy, the rotation rate of sunspot groups during the early stages of their evolution ($< 2-3$ days from their appearance) is about 2 percent greater than that of recurrent groups and  decreases steadily during their later evolution \citep[e.g., ][]{ZappalaZuccarello91}. However, we believe that it is unlikely that  a similar phenomenon may  explain the results obtained for CoRoT-Exo-4a because our active longitude drift lasts for several tens of days. In conclusion, surface differential rotation in combination  with active regions located at different latitudes seems to be the most likely explanation for the results presented in Sect.~\ref{results}. Although the observed active longitudes rotate faster than the planet, the existence of a latitude exactly synchronized to the planetary orbit cannot be excluded because we have no information on the latitude range covered by the active regions. } 

The { quasi-synchronous rotation} status of CoRoT-Exo-4a  might have been attained when the star, although already close to the ZAMS, was still magnetically locked to the inner parts of a circumstellar disc.  If the planet had already migrated to the radius where the disc was truncated by the stellar magnetic field 
\citep{RomanovaLovelace06},  the system started its evolution in a rotational and orbital configuration close to the present status. The subsequent tidal interaction between the star and the planet could not significantly change the angular momentum of the planet and the star, as noticed by \citet{Aigrainetal08}. The magnetic braking of the star after the dissipation of the disc is expected to be negligible given its mid-F spectral type and an initial rotation period of $\sim 9$ days 
\citep{WolffSimon97}.

The surface differential rotation found through our spot modelling is not remarkably dependent on the adopted value of the facular-to-spotted area ratio $Q$ and is at least $6-7$ times that derived for CoRoT-Exo-2a that is rotating in  $4.52$ days and is of spectral type G7V \citep{Lanzaetal08}. 
For  $\tau$ Bootis, having $T_{\rm eff} = 6360$~K and a rotation period of $\sim 3.3$ days, \citet{Catalaetal07} found a relative differential rotation  $\Delta \Omega / \Omega \approx 0.15-0.18$  between the equator and the pole by means of line profile modelling. This compares well with our value, given that our determination is only a lower limit. The same method was  applied by \citet{Reiners06} to a sample of F-type stars finding projected pole-equator relative amplitudes that are comparable to that of CoRoT-Exo-4a when one considers objects with a similar effective temperature and  $15 \leq v \sin i \leq 30$ km s$^{-1}$. Note that the small $v \sin i$ of our star prevents the application of line profile analysis methods to measure its surface differential rotation \citep{Reiners06}, therefore, 
a spot modelling approach is the only possibility to estimate its surface shear. 

Theoretical predictions for the amplitude of the surface differential rotation in main-sequence F-type stars have recently been published by \citet{KukerRuediger05}. Their model is in general agreement with the present findings although the lack of a precise theory for the interaction between stellar rotation and turbulent convection makes a detailed quantitative comparison premature.  

{The possibility that stellar activity is influenced by the  hot Jupiter in CoRoT-Exo-4a is suggested by
the present models, in particular by those assuming that the photometric modulation is produced only by cool spots (i.e., with $Q=0$). The plots in Fig.~\ref{long_mean} suggest that  the formation of spots may be triggered in an active region {located at the subplanetary longitude}, i.e., the emergence of magnetic flux may be promoted there in some way by the close planet. 
 It is important to note that tidal effects cannot account for this phenomenon because there is only one region of enhanced activity on the star during most of the time, and not two separated by $180^{\circ}$, as expected in that case.
The possibility that a hot Jupiter may affect in some way the dynamo action in the subsurface layers of the convection zone of its host star has been discussed by \citet{Lanza08}, and the present results give some support to his speculations.  
However, a word of caution is certainly appropriate in the present case  because the time interval covered by our data is rather limited. Therefore, the supposedly enhanced activity related to the planet might still be the result of statistical fluctuations in the appearance of small active regions over the star. Moreover, the relationship between enhanced  activity and subplanetary longitude appears to be less clear when we consider models with $Q > 0$, also  because the migration rates of the main active longitudes are greater.}

\section{Conclusions}
\label{conclusions}

Our analysis shows that the rotation of CoRoT-Exo-4a is { quasi-synchronized}, with the orbital motion of the planet. { Specifically, the active longitude having the greatest spotted area shows an angular velocity that is only $3-5$ percent greater than the orbital angular velocity. Assuming that it traces the angular velocity of stellar rotation at some latitude, this translates into an upper limit for the difference between stellar and orbital angular velocities at that latitude.} 
 Moreover, we find  evidence for a significant differential rotation of the star, with a lower limit $\Delta \Omega / \Omega = 0.057 \pm 0.015$, when the optimal facular-to-spotted area ratio $Q=4.5$ is adopted. 

It cannot be excluded that the planet may affect in some way the longitude distribution of the active regions on the photosphere of the star. { Specifically, there appears to be an enhacement of spot activity close to the subplanetary longitude, particularly for the models computed without faculae}. Unfortunately, the short duration of the present time series (only 58 days, i.e., 6.3 stellar rotations)  does not allow us to derive definite conclusions on such a kind of star-planet magnetic interaction. 

A long-term monitoring of the star from the ground may help to clarify this issue, e.g., through a measurement of the rotational modulation of the chromospheric Ca II H \& K flux that should be a good indicator of a possible star-planet interaction, as in the case of $\tau $ Boo or $\upsilon$ And \citep{Walkeretal08,Shkolniketal08}.

\begin{acknowledgements}
The authors are grateful to the Referee, Prof. Gordon Walker, for a careful reading of the manuscript and several valuable comments. 
This work has been partially supported by  the Italian Space Agency (ASI) under contract  ASI/INAF I/015/07/0,
work package 3170. Active star research and exoplanetary studies at INAF-Osservatorio Astrofisico di Catania and Dipartimento di Fisica e Astronomia dell'Universit\`a degli Studi di Catania 
 are funded by MIUR ({\it Ministero dell'Istruzione, dell'Universit\`a e della Ricerca}), and by {\it Regione Siciliana}, whose financial support is gratefully
acknowledged. 
This research has made use of the ADS-CDS databases, operated at the CDS, Strasbourg, France.
\end{acknowledgements}

\end{document}